\documentstyle[aps,epsf]{revtex}
\begin{document}
\wideabs{
\title{Simulations of Interacting Membranes}
\author{Nikolai Gouliaev and John F. Nagle}
\address{Department of Physics\protect\\
Carnegie Mellon University , Pittsburgh, PA 15213}
\date{\today}
\maketitle

\begin{abstract}
The liquid crystalline model biomembrane system consisting of a stack 
of interacting membranes is studied by the newly developed Fourier
Monte Carlo simulation technique. In comparison to perturbation
theory, substantial quantitative discrepancies are found that affect
determination of interbilayer interactions.
A harmonic theory is also routinely used
to interpret x-ray scattering line shapes; this is shown to be valid
because the distance dependence of the simulated correlation functions
can be fairly well fit by the harmonic theory.
\end{abstract}

\draft
}

\newpage

\begin{figure}[h]
\begin{center}
\leavevmode
\epsfxsize 8.5cm
\epsffile{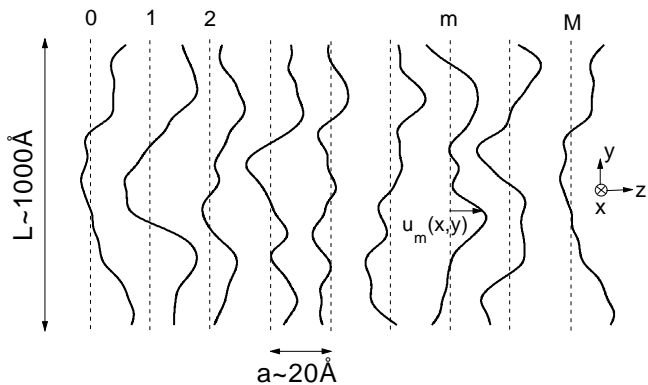}
\caption{Snapshot of a slice through a simulated stack of $M=8$ 
two-dimensional $L \times L$ fluctuating
membranes. Since internal membrane structure is irrelevant here, each
membrane is depicted as a line.  The average position of each membrane
is shown by a dashed line.}
\label{StackPicture}
\end{center}
\end{figure}

Stacks of lipid bilayers (see Fig. \ref{StackPicture}) are model
systems for biomembranes 
that are much studied for two reasons.  First, such stacks diffract
fairly well and this facilitates determination of the structure of
individual membranes, which is of primary interest in biophysics.
However, these stacks are not crystals with the long range order
that is assumed in traditional biophysical analysis, but smectic
liquid crystals with quasi-long-range order.
Therefore, quantitative use of the scattering intensity
for structure determination requires correction for the fluctuations 
endemic in such systems \cite{JN96}.
A harmonic theory \cite{Caille72,ZhangSuterNagle94} that predicts
power law tails in the scattering line shapes fits membrane data very well
\cite{Zhang96,Petrache98}, but the anharmonicities that are inherent
in realistic potentials have remained a concern for quantitative
interpretation \cite{Lem96,Zhang95}, even though a renormalization group
analysis suggested that such effects are small \cite{Grin81}.

Stacks of bilayers are also much studied because they
provide ideal environments to study fundamental interactions between
bilayers, especially since the range of interbilayer distances $a$ can
be systematically varied by applying osmotic pressure $P$
\cite{RandParsegian89}.  The corresponding theory \cite{PP92}
is an approximate first-order perturbation theory that again
relies on harmonic assumptions, such as the normality of the
probability distribution function (pdf) of the interbilayer spacing. 
While this theory has been a valuable guide, use of it to extract
fundamental interbilayer interactions from $P(a)$ data may be inaccurate.

Both these issues are addressed using Monte Carlo simulations
with realistic intermembrane potentials for the biologically
relevant regime where the interbilayer water spacing $a$ is of
order $5-30\AA$ and each membrane is flexible with bending modulus
$K_c{\simeq}10-25 k_B T$.  In this regime, sometimes called the soft
confinement regime \cite{PP92}, it is usually supposed that the
primary interbilayer interactions for dipolar lipids are the
attractive van der Waals potential and the repulsive hydration potential,
\begin{equation}
V(z) = - \frac{H}{12{\pi}z^2} + A\lambda e^{-z/\lambda} ,
\label{V(z)}
\end{equation}
where $z$ is the local distance between two membranes\cite{FN1}.
These interactions are significantly anharmonic, to the extent that
the potential of a membrane midway between two
neighboring membranes (at the dashed positions in Fig. \ref{StackPicture})
may have a maximum instead of a minimum.
The contrasting regime, sometimes called the hard confinement regime,
consists of only excluded volume or steric interactions between
neighboring membranes.  That regime is appropriate when 
$a$ is of order $100\AA$ because the hydration force is short range
($\lambda{\approx}2 \AA$).
For hard confinement the effective interbilayer force is the entropic
fluctuation pressure which decays as ${a}^{-3}$ \cite{Helfrich78}.
Fluctuation forces also exist in the soft confinement regime,
and are determined by our simulations.

In addition to the interbilayer interactions in Eq.(\ref{V(z)}),
the energy of each membrane includes a bending term, proportional
to the square of the local curvature of the membrane. 
Let $u_{m}(x,y)$ be the local displacement of the $m$th membrane from
its average position as shown on Fig. \ref{StackPicture}.  Periodic
boundary conditions are imposed 
in the plane of each membrane and also along the stack, so that
$u_{M}{\equiv}u_0$. The membranes can collide,
but cannot overlap, so that $u_{m+1}+a {\geq} u_{m}$, where $a$ is the
average distance between membranes. The Hamiltonian of the stack is then: 
\begin{equation}
H {\!}={\!} \sum_{m=0}^{M-1} \int \left[
\frac{K_c}{2} ({\Delta}u_m)^2 {\!}+{\!} V(u_{m+1}{\!}+{\!}a{\!}-{\!}u_m)
\right]dxdy
\label{MultipleH}
\end{equation}

The simulation method, called the Fourier Monte Carlo method,
was developed for single membranes between hard
walls \cite{NG_JN_97} and is easily extended to stacks of membranes.
Each membrane in the stack is represented by
a complex array of dimensions $N \times N$ of Fourier displacement amplitudes.
Instead of moving one lattice site at a time, moves
are made in Fourier space and a whole membrane is displaced in each move.
This allows larger moves and faster equilibration,
without incurring large increases in the bending energy.
One difference with our previous simulations \cite{NG_JN_97} is that 
a fixed osmotic pressure $P$ ensemble is employed instead of
the previous fixed $a$ ensemble, so that $a$ is obtained
as a function of $P$ rather than {\it vice versa}.
Of course, use of the $P$ ensemble is fundamentally no different, but
it does have better convergence properties that we now discuss.

\begin{figure}[h]
\begin{center}
\leavevmode
\epsfxsize 8.5cm
\epsffile{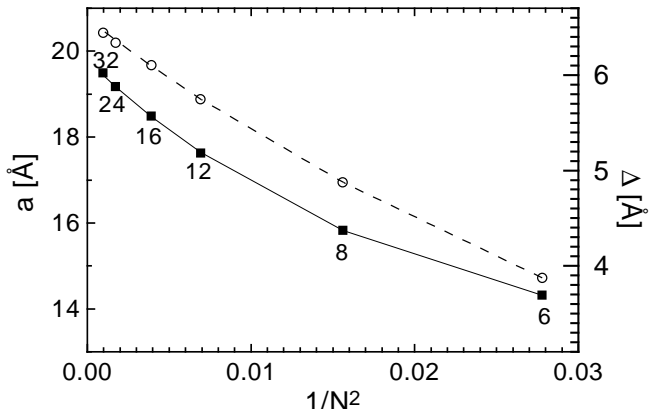}
\caption{Effect of finite density ($N=6,...,32$) 
on $a$(solid squares) and $\Delta$(open circles and right hand
ordinate) for realistic interaction parameters given in \protect\cite{FN2} 
and for $P=10^4 erg/cm^3$.} 
\label{hpv6scal_d_delta}
\end{center}
\end{figure}
Simulations performed systematically as a function of lattice size,
density of lattice points, and number of membranes in the stack
show that accurate results for infinite, continuous membranes in
infinite stacks can be obtained at one $P$ in real time of the order of
one day on a Pentium Pro PC.  The most sensitive finite-size parameter is
the ``density'' of each membrane $N/L$,
since when $N$ is varied from 6 to 32, the root mean square fluctuation in
nearest neighbor spacing, defined as $\Delta$, can easily change by 40\%,
and the changes in $a$ are also significant.   This is shown in
Fig.\ref{hpv6scal_d_delta}, which also shows that accurate
values can be obtained by extrapolation.  By comparison, variations
with lateral system size $L$ at fixed $N/L$ are negligible (${\approx}0.2\%$
for $L{\geq}700\AA$), as are variations with $M$ (${\approx}1\%$ for
$M{\geq}8$) for $a$ and $\Delta$.

It may be noted that stacks of several membranes ($M{\approx}4$) have been
previously considered \cite{GompperKroll89,NetzLipowsky93,NetzLipowsky95}, 
but mostly for the critical phenomenon of unbinding; this occurs in the
limit of large average membrane spacing, where the van der Waals
interaction is the main one, in addition to the spatial constraints.
We have also performed simulations in the hard confinement regime and
obtained results for the Helfrich fluctuation free energy 
$c_{fl}T^2/K_ca^2$ with $c_{fl}{\approx}0.1$, in agreement
with \cite{GompperKroll89,NetzLipowsky95}.

\begin{figure}[h]
\begin{center}
\leavevmode
\epsfxsize 8.5cm
\epsffile{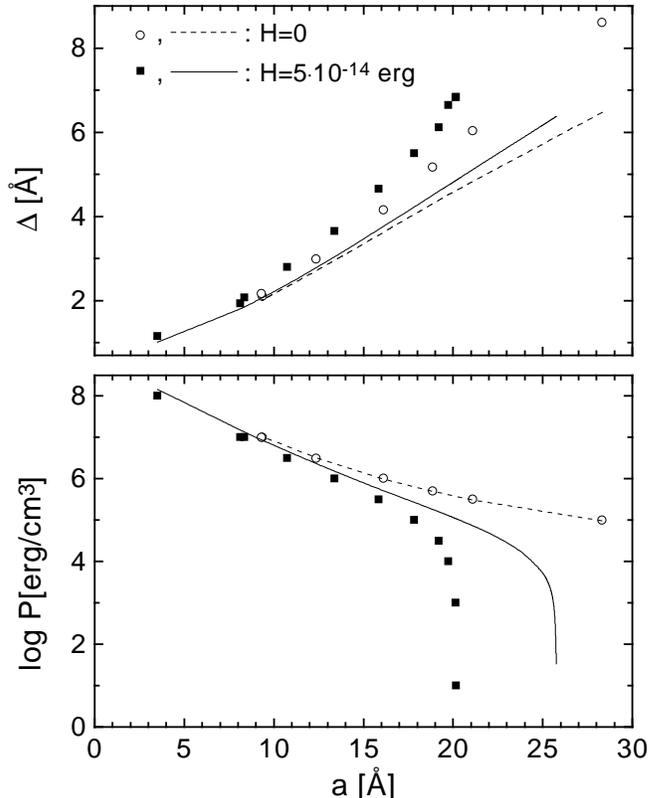}
\caption{Simulation results (symbols) and perturbation theory (lines)
for $\Delta$ and ${\log}P$ versus $a$ for the parameter set in
\protect\cite{FN2} with attractive interaction (solid squares and
lines) and with no attractive interaction (open circles and dashed lines).}
\label{hpv6and1_delta_logp}
\end{center}
\end{figure}
Figure \ref{hpv6and1_delta_logp} shows results for $P$ and ${\Delta}$
as functions of $a$ for realistic values of the potentials \cite{FN2}.
Also shown are results for a simpler case when the attractive van der
Waals force is absent so that the potential experienced by each
membrane has a minimum in the middle between its neighbors and is 
therefore more like a harmonic potential. Fig.\ref{hpv6and1_delta_logp}
also shows results based on the perturbation approximation\cite{PP92} 
which was developed for a single membrane between hard walls.
After comparing to harmonic theory for multiple
membranes \cite{Petrache98}, we adjusted \cite{PP92} for the case of
multiple membranes by putting a factor of
$4/\pi$ into the relations for the fluctuational free energy and for
$\Delta^2$, and then followed essentially the same procedure as in
\cite{PP92}. With this factor, agreement between the perturbation theory
and the simulations is quite good for small $a$ where the total interaction
is harmonic-like because the repulsive hydration potential is dominant
and the fluctuations are relatively small.
The simulations and the theory also agree for $P(a)$ when there is no
attractive van der Waals interaction.  However, at larger $a$, the
simulated $\Delta$ increases with $a$ faster than the perturbation 
approximation for either intermembrane potential.  
A large difference between the simulation results and the perturbation
theory occurs when the potential has an attractive van der Waals part.
The theory predicts that $a_o=25.8\AA$ for $P=0$, more than
$5\AA$ larger than the true value $a_o=20.2{\pm}0.1\AA$ obtained by
simulation. It is also of interest to compare the values of $\mu
\equiv (\Delta/a)^2$ to the hard confinement values.  In
Fig.\ref{hpv6and1_delta_logp} the range of $\mu$ is $0.06-0.12$,
in good agreement with experiment \cite{Petrache98}, but
considerably smaller than the hard confinement estimates $0.16-0.21$
\cite{PP92,Helfrich78,NetzLipowsky95,FN3}.

\begin{figure}[h]
\begin{center}
\leavevmode
\epsfxsize 8.5cm
\epsffile{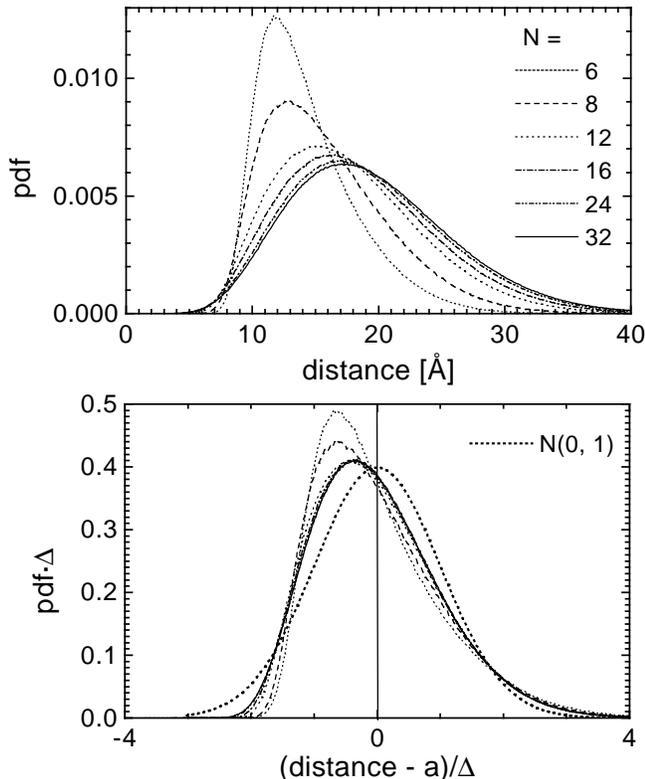}
\caption{Distribution of the nearest neighbor distance for parameters
in \protect\cite{FN2} and $P=10^4erg/cm^3$ for different membrane
densities. The bottom figure
shows the same results as the top one, scaled and shifted so as to
facilitate comparison to the standard normal distribution, denoted
$N(0,1)$.}
\label{hpv6hist_pdf}
\end{center}
\end{figure}
Existing approximations\cite{PP92,SO86} assume that the distribution of
distance between two nearest neighbor membranes is essentially Gaussian.
Simulation results in Fig. \ref{hpv6hist_pdf} demonstrate that the 
actual pdf differs substantially from a normal distribution.
Although the distribution is more non-normal for small $N$, it is
nevertheless clear that even the limiting distribution is not normal.
A distinguishing feature of the true pdf is a rapid decay at small
distances. This validates the use of a cutoff \cite{FN1} near $2\AA$ to
avoid the formal divergence of the vdW potential. Another feature
evident in Fig. \ref{hpv6hist_pdf} is the asymmetry of the actual pdf 
which shows that large fluctuations to larger intermembrane spacings are
more probable than large fluctuations to smaller spacings.
Overall, the shape of pdf is consistent with that of the interaction potential.

\begin{figure}[h]
\begin{center}
\leavevmode
\epsfxsize 8.5cm
\epsffile{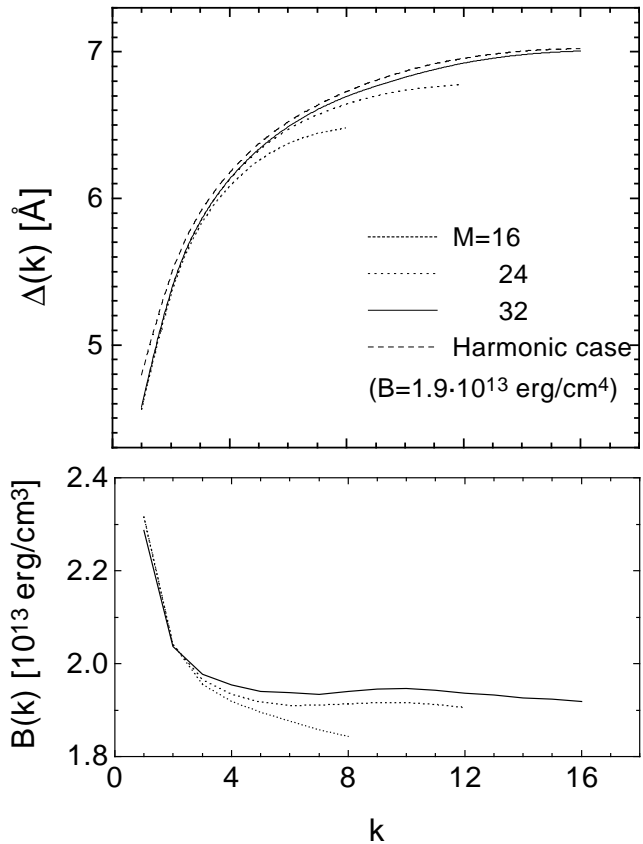}
\caption{Root mean square fluctuations $\Delta(k)$ between $k$th neighbor
membranes for a stack with different numbers $M$ of membranes and for
the parameter set in \protect\cite{FN2} (except that $L=1400\AA$)
at $P=3.16{\cdot}10^5 erg/cm^3$. Also shown is $\Delta(k)$,
exactly computed for the case of harmonic interactions with
compression modulus $B=1.9{\cdot}10^{13}erg/cm^4$. The bottom figure
shows the effective harmonic compression modulus $B(k)$ for each $k$ and
$M$.} 
\label{delta_k6_delta_b}
\end{center}
\end{figure}

We turn next to the issue of whether harmonic fluctuation
theory \cite{Caille72,ZhangSuterNagle94} should be expected to be
reliable when interpreting detailed line shapes
\cite{Zhang96,Petrache98} from stacks of bilayers and smectic
liquid crystals that have strongly anharmonic interactions.
The most important quantity for determining the line shapes
for powder samples \cite{ZhangSuterNagle94} is the correlation function in
the $z$-direction, which is essentially the $k$ dependence of
the mean square relative displacement of two membranes of the stack
$\Delta^2(k)\equiv{\langle}(u(m+k)-u(m))^2{\rangle}$, where $m$ and $m+k$
are the membrane indices, and averages over $m$ are performed
for simulation efficiency.  
Fig.\ref{delta_k6_delta_b} shows profiles of $\Delta(k)$,
obtained for stacks with various numbers $M$ of membranes. 
Convergence with $M$ suggests that values of ${\Delta(k)}$ are
sufficiently accurate for $k<M/4$.  However, to minimize the
finite size effect in a comparison with harmonic theory, 
Fig. \ref{delta_k6_delta_b} compares the results of the simulation
with $M=32$ with the exact harmonic result, also for $M=32$.
In the harmonic theory the bare interbilayer interactions are
approximated with a compression modulus $B$. In
Fig. \ref{delta_k6_delta_b} a value of $B=1.9{\cdot}10^{13}erg/cm^4$  
was chosen to match the large $k$ end of the $M=32$ curve.
The resulting $\Delta(k)$ profile allows one to see that
$\Delta{\equiv}\Delta(1)$ is in fact a good proxy for describing the
long-range correlations, since the difference between $\Delta$,
implied by the ``harmonic curve'', and the actual $\Delta$ for the
stack is about only $0.2\AA$, i.e. relatively small compared to
${\Delta}=4.6\AA$. Another way to see how interactions are
renormalized from short to long range is to compute, for different
$k$, the implied strength $B(k)$ of the harmonic potential that would
result in the same value of $\Delta(k)$ as obtained by simulations
for a stack with
realistic interactions. The bottom panel of Fig.\ref{delta_k6_delta_b}
presents a plot of the harmonic value of $B(k)$ required to
give the simulation value of ${\Delta}(k)$.  This shows that, for
large $k$, the system can be reasonably well approximated by one
with harmonic interactions with constant $B$.

How is it that the harmonic theory works quite well for the
correlation functions in the preceding paragraph and not so well 
for $P$ and $\Delta$ in Fig. \ref{hpv6and1_delta_logp}?
The answer is that the perturbation theory does not yield
the best value of $B$; for the example in Fig.\ref{delta_k6_delta_b}
the theory yields a larger $B=5.4{\cdot}10^{13} erg/cm^4$ which
accounts for the smaller value of $\Delta$ in
Fig. \ref{hpv6and1_delta_logp}.  

\begin{figure}[h]
\begin{center}
\leavevmode
\epsfxsize 8.5cm
\epsffile{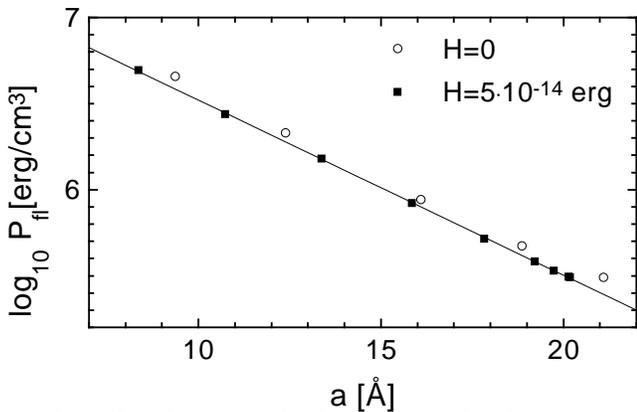}
\caption{Simulation results for $P_{fl}$ vs. $a$ for the parameter set
in \protect\cite{FN2} and also for $H=0$. The slope is
$\lambda_{fl}=4.34 \AA$.}
\label{hpv6_pfluct}
\end{center}
\end{figure}

We turn finally to the entropic fluctuation pressure in a stack of membranes,
which is defined to be the difference between the applied pressure,
and the pressure due to the bare van der Waals and hydration
interactions.  Perturbation theory \cite{PP92}, experiment
\cite{Petrache98} and simulations on a single membrane between hard
walls \cite{NG_JN_97} all  agree that the decay of the fluctuation
pressure is closer to exponential, with a decay length
${\lambda}_{fl}$, although the value of ${\lambda}_{fl}$ found in both
experiment and simulations is larger than the perturbation theory
prediction ${\lambda}_{fl}=2{\lambda}$. Fig. \ref{hpv6_pfluct} shows
the same result for simulations of stacks.
\vspace{-0.3in}

A long range goal is to obtain values of the interbilayer interaction
parameters; the traditional analysis \cite{RandParsegian89} uses
osmotic pressure $P(a)$ data, which has recently been supplemented
by fluctuation ${\Delta}(a)$ data \cite{Petrache98}.
One of the main results of this paper indicates that the $\Delta$ data are 
indeed valid, even though the analysis of the basic x-ray scattering
data is based on a harmonic theory.  However, the intrinsic anharmonic
nature of realistic interactions between bilayers in stacks makes it
difficult to devise quantitatively accurate analytic or perturbation
theories.  We show here that the Fourier Monte Carlo method is
sufficiently fast that it provides a viable alternative.  Indeed, it
is now possible to consider using it as part of a comprehensive data
analysis program to determine the best values of the fundamental
interaction parameters.

Acknowledgments: We thank Horia Petrache for many useful
discussions and especially for his insight that at large
distances, smaller deviations from harmonicity may be expected.
This research was supported by the U. S. National
Institutes of Health Grant GM44976.


\begin{thebibliography}{99}
\bibitem{FN1}
The formal divergence of the van der Waals potential at $z=0$ is
non-physical, being masked by stronger steric interactions.  In the
simulations, a constant potential is used when $|z|$ is smaller than
where $V(z)$ has a maximum.  For these values of $z$, which are typically
less than $2\AA$ the hydration potential is so large that few membrane
pairs approach this closely.
\bibitem{FN2}
In most parts of this Letter we use the following parameters:
$A{\!}{\!}={\!}{\!}10^9 erg/cm^3$; $H{\!}={\!}5{\cdot}10^{-14} erg$;
$K_c{\!}={\!}0.5{\cdot}10^{-12} erg$; 
$\lambda{\!}={\!}1.8\AA$; $T{\!}={\!}323K$; $L{\!}={\!}700\AA$;
$M{\!}={\!}8$.
\bibitem{FN3}
Our simulations for hard confinement yield ${\mu}=0.16$.
\end{thebibliography}
\end{document}